\documentclass[apj]{emulateapj}
\usepackage{ulem}
\usepackage{color}

\newcommand{\icarus}{Icarus}

\shorttitle{CAPTURE INTO MEAN MOTION RESONANCES}

\shortauthors{Ogihara et al.}

\begin{document}

\title{CONDITION FOR CAPTURE INTO FIRST--ORDER MEAN MOTION RESONANCES
AND APPLICATION TO CONSTRAINTS ON ORIGIN OF RESONANT SYSTEMS}

\author{Masahiro Ogihara}
\affil{Nagoya University,
Furo--cho, Chikusa--ku, Nagoya, Aichi 464-8602, Japan}
\email{ogihara@nagoya-u.jp}

\and
\author{Hiroshi Kobayashi}
\affil{Nagoya University,
Furo--cho, Chikusa--ku, Nagoya, Aichi 464-8602, Japan}
\email{hkobayas@nagoya-u.jp}

%\begin{document}
\begin{abstract}
We investigate the condition for capture into first--order mean motion resonances
using numerical simulations with a wide range of various parameters. 
In particular, we focus on deriving the critical migration timescale for capture into
the 2:1 resonance; additional numerical experiments for closely spaced resonances
(e.g., 3:2) are also performed. We find that the critical migration timescale is 
determined by the planet--to--stellar mass ratio, and its dependence exhibits 
power--law behavior with index $-4/3$. This dependence is also supported by
simple analytic arguments. We also find that the critical migration timescale
for systems with equal--mass bodies is shorter than that in the restricted problem;
for instance, for the 2:1 resonance between two equal--mass bodies, the critical 
timescale decreases by a factor of 10.
In addition, using the obtained formula, the origin of observed systems that include first--order
commensurabilities is constrained. 
Assuming that pairs of planets originally form
well separated from each other and then undergo convergent migration and are captured in 
resonances, it is possible that a number of exoplanets experienced rapid orbital migration.
For systems in closely spaced resonances, the differential migration timescale between
the resonant pair can be constrained well; it is further suggested that 
several exoplanets underwent migration that can equal or even exceed the type I
migration rate predicted by the linear theory. This implies that some of them may
have formed {\it in situ}. 
Future observations and the use of our model will allow us to statistically determine 
the typical migration speed in a protoplanetary disk.

\end{abstract}
\keywords{methods: numerical -- celestial mechanics -- planets and satellites: formation
-- planet--disk interactions}

\clearpage

\section{INTRODUCTION}
When two bodies are in a mean motion resonance, their orbital periods are close to
a ratio of two integers, which stabilizes the system in many cases.
The Laplace 4:2:1 resonance among the Galilean satellites and the 3:2 resonances 
between Neptune and trans--Neptunian objects 
(TNOs) are well--known examples in the solar system.

More than 30 exoplanet systems also include
confirmed planets exhibiting mean motion resonances. In addition, \textit{Kepler} has 
detected a large number of planet pairs near mean motion resonances 
(e.g., \citealt{lissauer_etal11}; \citealt{baruteau_papaloizou13}), although they are not
necessarily in resonances.
Many exoplanets are not near mean motion resonances
(e.g., \citealt{mayor_etal09}). However, according to \citet{ogihara_ida09}, such non resonant
configurations can be established through orbit crossing among the planets after they are
captured in mutual mean motion resonances. Thus, mean motion resonances 
may have played an important role in the planet formation process.

To date, mean motion resonances have been studied from many perspectives
(e.g., \citealt{goldreich65}; \citealt{wisdom80}; \citealt{henrard82};
\citealt{weidenschilling_davis85}; \citealt{nelson_papaloizou02};
\citealt{kley_etal04}; \citealt{terquem_papaloizou07}; \citealt{raymond_etal08};
\citealt{ogihara_ida09}; \citealt{ogihara_etal10}).
Resonant configurations are thought to arise primarily from convergent migration
(e.g., \citealt{snellgrove_etal01}).
Several efforts have been made to derive
a critical differential (relative) migration timescale $t_{a,{\rm crit}}$
above which mean motion resonances can be formed. 	

\citet{friedland01} analytically studied the restricted three--body
problem with adiabatic migrating bodies, where the relative migration timescale is much
longer than the resonant libration timescale. In contrast, recent studies have adopted numerical
methods. One approach is to perform direct \textit{N}--body calculations (e.g., 
\citealt{ida_etal00}; \citealt{wyatt03}), which take a direct summation of the mutual interaction between the bodies.
Another approach uses the Hamiltonian model (e.g., 
\citealt{quillen06}; \citealt{mustill_wyatt11}),
where the canonical equations of the Hamiltonian are solved numerically with some approximations.
These studies examine the dependence of the critical migration timescale on the mass but
consider the restricted three--body problem with a massive planet and a massless test particle.
\citet{rein_etal10,rein_etal12} performed direct \textit{N}--body calculations with equal--mass
planets. However, because their goal was to specify the origin of individual systems, 
the planetary mass was not usually treated as a parameter.

Our main aim is to derive an empirical formula for the critical migration timescale of 
capture into first--order $p+1:p$ mean motion resonances by performing direct \textit{N}--body simulations.
We handle the physical variables (e.g., mass and damping timescales) as parameters and
vary them over wide ranges. In this way, the dependences of the critical migration
timescale on the parameters can be obtained.
Although the equal--mass case is of primary importance for the study of observed exoplanets in resonances,
previous general studies, which are not restricted to particular systems, have not considered this situation; therefore, we examine the case
of equal--mass bodies. Captures into 2:1 resonances, which are the outermost first--order 
resonances, are extensively studied in this paper; in addition,
more closely spaced commensurabilities (e.g., 3:2 and 4:3) are also examined.
From the empirical formula based on our numerical results, we try to constrain the origin of the orbital architecture
of planetary systems exhibiting commensurabilities.

The structure of this paper is as follows. 
In Section~\ref{sec:model}, we describe the numerical methods; in Section~\ref{sec:results},
we present and summarize the results of \textit{N}--body simulations. In Section~\ref{sec:discussion},
we derive the mass dependence of the critical migration time, and
in Section~\ref{sec:compare}, we compare our results with those of previous
studies.
In Section~\ref{sec:constraints}, we apply these results to
systems with resonances and discuss their origin, and in Section~\ref{sec:conclusions},
we offer our conclusion.

\section{NUMERICAL MODEL}
\label{sec:model}
\begin{figure}[htbp]
\epsscale{1}
\plotone{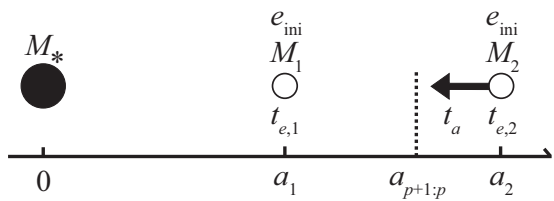}
\caption{The model considered in this paper. Two bodies of masses $M_1$ and $M_2$ with
orbits of semimajor axes $a_1$ and $a_2$, respectively, around a central body with mass $M_*$. 
The initial eccentricities are set to $e_{\rm ini}$. The orbits of the inner and outer bodies
are circularized on timescales of $t_{e,1}$ and $t_{e,2}$, respectively. We assume
the outer body undergoes inward migration on the timescale of $t_a$ to see capture into a
$p+1:p$ mean motion resonance.}
\label{fig:model}
\end{figure}

Figure~\ref{fig:model} presents the calculation model considered in this work.
Initially, two planets with masses of $M_1$ (inner body) and $M_2$ (outer body) 
are placed at semimajor axes $a_1$ and $a_2$, respectively. 
The initial eccentricity of these bodies is set to $e_{\rm ini}$, and their inclination
is also set to about $e_{\rm ini}$ in radian. Damping forces are applied to the bodies that
damp the eccentricity and semimajor axis on timescales of 
$t_e$ and $t_a$, respectively. We apply the $a$--damping force only to
one body (mostly to the outer body); therefore, $t_a$ is interpreted as the 
timescale of differential
migration between the bodies. The initial locations are adopted to be
separated from the resonance $a_{2}/a_{1} = [(p+1)/p]^{2/3}$; namely, $a_2 = 1.8 a_1$ for
2:1 resonance, and for other closer resonances (e.g., $p=2$) the orbits are set just inside 
the $p:p-1$ resonance $a_{2}/a_{1} = [p/(p-1)]^{2/3}$.
Thus, when $t_e$ is much shorter than $t_a$, the eccentricity is negligibly small at the 
resonant encounter.

The equation of motion for planet 2 is
\begin{eqnarray}
\frac{d^2 \textbf{\textit{r}}_2}{dt^2}
& = & -GM_* \frac{\textbf{\textit{r}}_2}{ |\textbf{\textit{r}}_2|^3} 
- GM_1
\frac{\textbf{\textit{r}}_2 - \textbf{\textit{r}}_1}{|\textbf{\textit{r}}_2 - \textbf{\textit{r}}_1|^3} 
\nonumber\\
&   &
- GM_1 \frac{\textbf{\textit{r}}_1}{ |\textbf{\textit{r}}_1|^3} 
- GM_2 \frac{\textbf{\textit{r}}_2}{ |\textbf{\textit{r}}_2|^3} 
\nonumber\\
&   &
+ \textbf{\textit{F}}_e + \textbf{\textit{F}}_a,
\end{eqnarray}
where the first term on the right--hand side is the gravitational force of the central star, 
the second term is the mutual gravity between the bodies, and the third and fourth are 
the indirect terms. 
$\textbf{\textit{F}}_e = - 2 (\dot \textbf{\textit{r}}_2 \cdot
\textit{\textbf{r}}_2)\textit{\textbf{r}}_2/|\textit{\textbf{r}}_2|^2 t_e$ and
$\textbf{\textit{F}}_a = - \dot \textit{\textbf{r}}_2/t_a$ are the specific forces for $e$--damping
and $a$--damping, respectively (see, for example, \citealt{mcneil_etal05}). 
The orbits are directly integrated with the fourth--order Hermite scheme
\citep{makino_aarseth92} over about 0.2 times the migration timescale, which means that
although the long--term stability is not exmined in this study, a temporary capture with a
capture duration of less than about 0.1 times the migration timescale can be excluded.

In order to derive critical values of $t_a$ for resonance capture,
numerical simuations are performed with different $t_a$. The sampling interval of $t_a$
adopted in this study is usually 0.2 in a logarithmic scale.
We treat
the masses ($M_1$ and $M_2$), the $e$--damping timescales ($t_{e,1}$ and $t_{e,2}$), 
and the initial eccentricity ($e_{\rm ini}$) as
parameters. For a fiducial model, the values are $M_1=1 M_\oplus$, $M_2=10^{-2} M_\oplus$, 
$t_{e,1}=\infty$, $t_{e,2}=10^3 T_{\rm K}$, and $e_{\rm ini}=10^{-4}$,
as shown in Table~\ref{tbl:parameters}. Here, $T_{\rm K}$ indicates the
orbital period of the inner body.
We perform a series of calculations for a wide range of parameters to examine the 
dependences of resonant capture.

\begin{deluxetable}{cclcl}
\tablecolumns{8}
\tablewidth{0pc}
\tablecaption{Parameters}
\startdata
\hline \hline
Parameters & Fiducial value & \multicolumn{3}{c}{Parameter range}\\
\hline
$M_1 (M_\oplus)$ & 1 & $~10^{-1}$&--&$10^3$\\
$M_2/M_1$ & $10^{-2}$ & $~10^{-3}$&--&$1$\\
$t_{e,1} (T_{\rm K})$ & $\infty$ & $~10^2$&--&$\infty$\\
$t_{e,2} (T_{\rm K})$ & $10^3 $& $~10^2$&--&$\infty$\\
$e_{\rm ini}$ & $10^{-4}$ & $~10^{-4}$&--&$10^{-1}$
\enddata
\tablecomments{Parameters and their ranges assumed. 
Parameters are described in Figure~\ref{fig:model} and its caption.
}
\label{tbl:parameters}
\end{deluxetable}

The equation of motion can be normalized by the unit length $r_0$,
unit mass $M_*$, and unit time $\Omega_{\rm K}^{-1} = (GM_*/r_0^3)^{-1/2}$ as
\begin{eqnarray}
\frac{d^2 \tilde{\textbf{\textit{r}}}_2}{d\tilde{t}^2}
& = & - \frac{\tilde{\textbf{\textit{r}}}_2}{ |\tilde{\textbf{\textit{r}}}_2|^3} 
- \tilde{M_1}
\frac{\tilde{\textbf{\textit{r}}}_2 - \tilde{\textbf{\textit{r}}}_1}
{|\tilde{\textbf{\textit{r}}}_2 - \tilde{\textbf{\textit{r}}}_1|^3} 
\nonumber\\
&   &
- \tilde{M_1} \frac{\tilde{\textbf{\textit{r}}}_1}{ |\tilde{\textbf{\textit{r}}}_1|^3} 
- \tilde{M_2} \frac{\tilde{\textbf{\textit{r}}}_2}{ |\tilde{\textbf{\textit{r}}}_2|^3} 
\nonumber\\
&   &
+ \tilde{\textbf{\textit{F}}}_e + \tilde{\textbf{\textit{F}}}_a,
\end{eqnarray}
where values with tildes on top are those scaled by the unit values.
Therefore, by changing $M_1$ and $M_2$, we can discuss the dependence on
the stellar mass. Hereafter, we usually use $M_* = M_\odot$  and present
$M_1$ and $M_2$ in units of Earth mass for simplicity.
In this paper, we mostly present results assuming $M_1 \geq M_2$ because
we find that the condition for capture into resonance is approximately expressed by 
the mass of the larger body, which will be shown in Section~\ref{sec:internal_resonance}.

\section{RESULTS}
\label{sec:results}
\subsection{Procedure for Obtaining Critical Migration Timescale}
\begin{figure}[htbp]
\epsscale{1}
\plotone{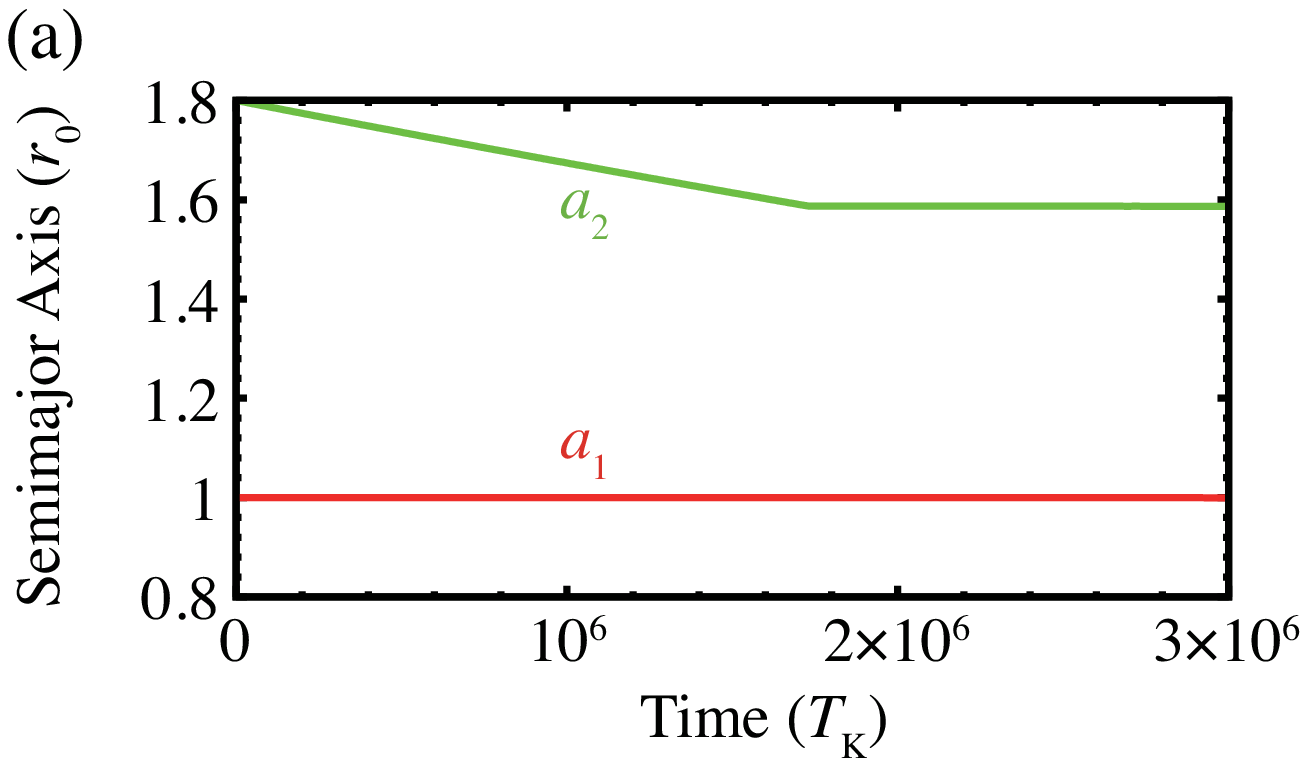}
\plotone{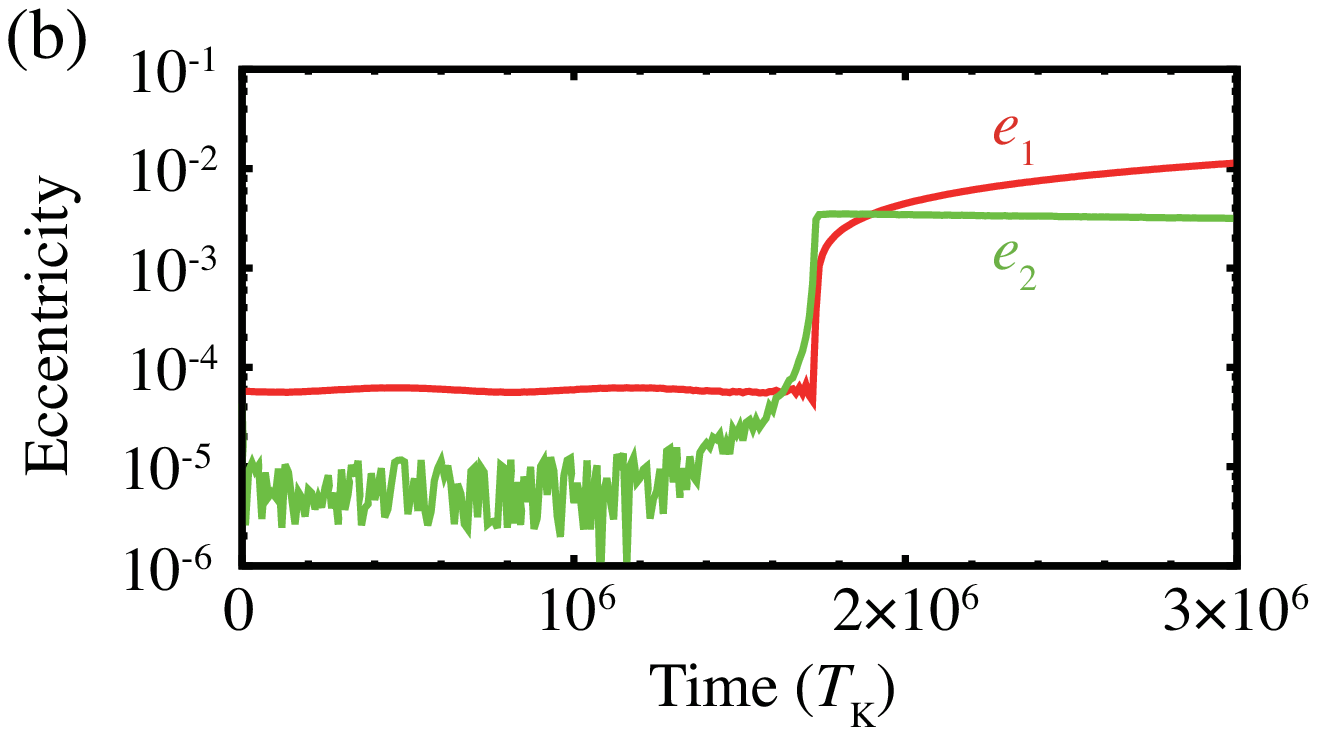}
\caption{Example of orbital evolution, where the migration timescale of the
outer body is assumed to be $1.59 \times 10^7 T_{\rm K}$. (a) Evolution of 
semimajor axes. (b) Evolution of eccentricities. Two bodies are captured into
the 2:1 mean motion resonance at $t \simeq 1.7 \times 10^6 T_{\rm K}$.}
\label{fig:t_a}
\end{figure}

We estimate the critical migration timescale in a manner similar to that of \citet{wyatt03}.
Orbital integrations of two planets are performed with various parameters to 
determine whether the planets are captured into 2:1 mean motion resonances.
Figure~\ref{fig:t_a} shows the evolution of the semimajor axes and eccentricities of the two 
planets with a migration timescale $t_a$ of $1.59 \times 10^7 T_{\rm K}$ in the 
fiducial model. When a 2:1 commensurability is formed at 
$t \simeq 1.7 \times 10^6 T_{\rm K}$, the eccentricities are excited.
We also confirm that the two resonant angles\footnote{The corresponding 
resonant angles are $\theta_1 = \lambda_1 - 2 \lambda_2 +\varpi_1$ and 
$\theta_2 = \lambda_1 - 2\lambda_2 + \varpi_2$, where $\lambda_i$ and $\varpi_i$ are mean 
orbital longitudes and longitudes of the pericenter, respectively.}
librate about fixed values.

\begin{figure}[htbp]
\epsscale{1}
\plotone{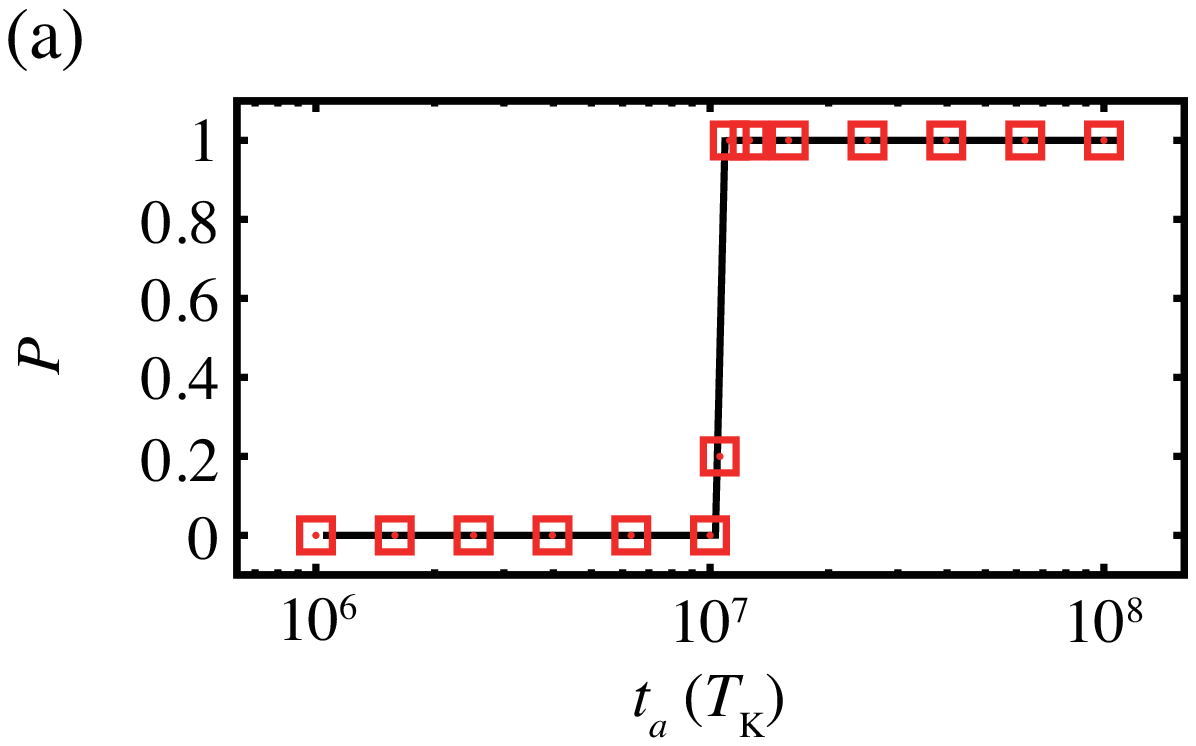}
\plotone{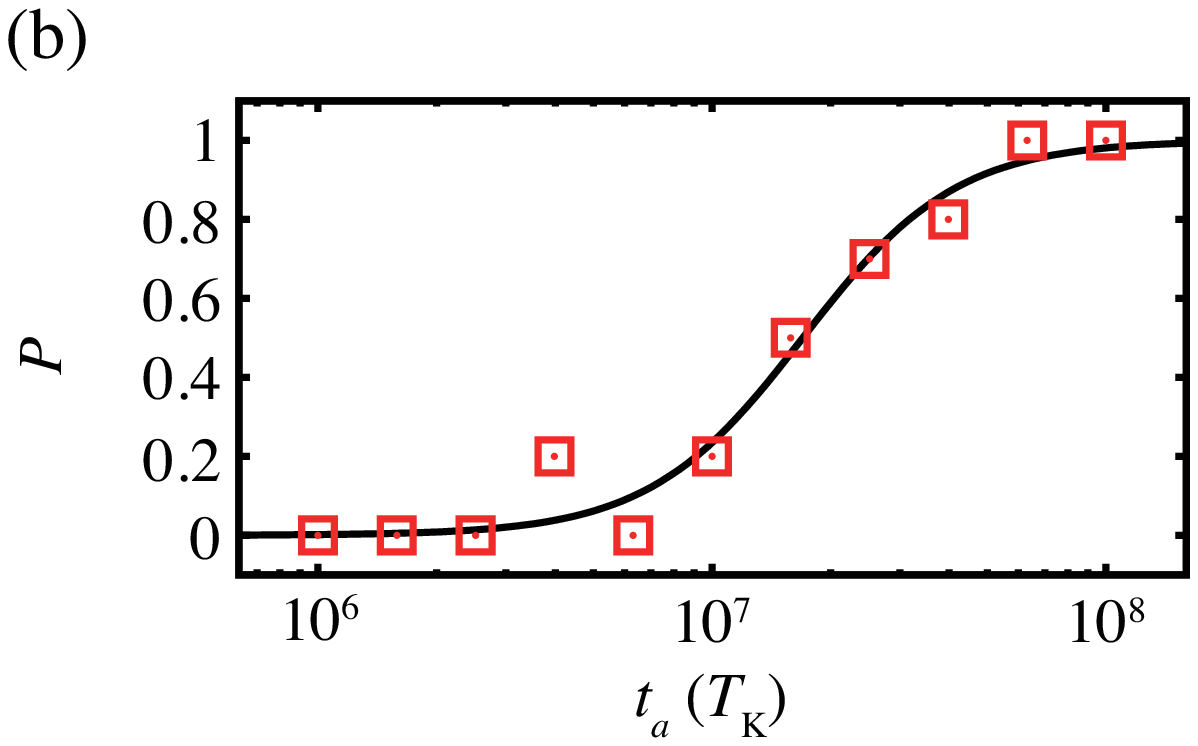}
\caption{Capture probability versus migration timescale for (a) the fiducial model and
(b) the model having relatively high initial eccentricities $(e_{\rm ini} \simeq 0.01)$.
A least-squares fit to the data (solid lines) allows the critical migration timescale ($t_a$ for $P=0.5$) to be specified.}
\label{fig:ta_P_fiducial}
\end{figure}

We made 10 runs for each migration timescale with different initial orbital phase angles
and derived the capture probability $P$ of 2:1 resonances . 
When the period ratio lies within an error of 1\%, we call it a resonant capture.
Note that the long--term stability of the resonant configuration is not considered.
Figure~\ref{fig:ta_P_fiducial}(a)
shows the capture probability for the fiducial model as a function of the migration time.
The probability $P$ is assumed to have the form
\begin{eqnarray}
P = \left[1 + \left(\frac{t_a}{t_{a,{\rm crit}}}\right)^\gamma \right]^{-1},
\end{eqnarray}
where $t_{a,{\rm crit}}$ is defined as the migration time for which
$P=0.5$ and $\gamma$ is a constant. 
Using a least--squares fit to the data (solid line), we specify $t_{a,{\rm crit}}$ and $\gamma$.
The capture probability increases sharply at $t_a \simeq 1.1 \times 10^7 T_{\rm K}$,
which means that the critical migration time $t_{a,{\rm crit}}$ can be estimated without
introducing large errors $(\gamma \la -300)$. 
Figure~\ref{fig:ta_P_fiducial}(b) presents the results for an initial eccentricity of
$e_{\rm ini} = 0.01$. 
In this case, the eccentricity is 0.01 at the resonant encounter because the inner body
does not undergo eccentricity damping. 
In contrast to low $e_{\rm ini}$, high eccentricities reduce $P$ at $t_a = 10^7$ -- $10^8 T_{\rm K}$, 
and $P$ gradually increases with $t_a$ $(\gamma \simeq -2.2)$.
This broadening of the capture probability curve at higher eccentricity is also
seen in previous studies (e.g., \citealt{quillen06}; \citealt{mustill_wyatt11}).
We find that the critical migration timescale can be sharply defined except when
the eccentricity is not small $(e \ga 0.01)$ at the resonant encounter. In the following
subsections, the value of $\gamma$ is explicitly described only when the transition
is not sharp enough.
The eccentricity dependence of $P$ is summarized in Section~\ref{sec:eini}.
We performed $\sim100$ runs for each set of parameters ($\sim15,000$ runs in total).

\subsection{Dependence on $M_1$}
\begin{figure}[htbp]
\epsscale{0.9}
\plotone{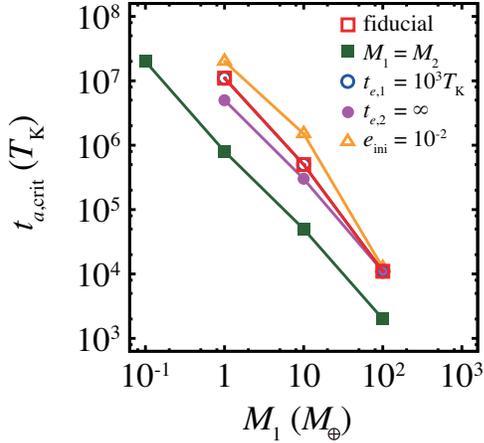}
\caption{Critical migration timescale versus mass for larger body. Open squares represent the
fiducial model. Filled squares, open circles, filled circles, and open triangles are the results for
$M_1/M_2=1, t_{e,1}/T_{\rm K}=10^3, t_{e,2}/T_{\rm K}=\infty$, and $e_{\rm ini}=10^{-2}$, 
respectively.
}
\label{fig:m1_ta}
\end{figure}

We first examine the dependence of the critical migration timescale on
the mass of the larger body $M_1$. Figure~\ref{fig:m1_ta} illustrates how $t_{a,{\rm crit}}$
varies with $M_1$. Open squares (connected by a solid line) represent the 
fiducial case, where $M_2/M_1 = 10^{-2}, t_{e,1}/T_{\rm K}=\infty, t_{e,2}/T_{\rm K}=10^3$, 
and $e_{\rm ini}=10^{-4},$ with different values of $M_1$. Filled squares, open circles, filled circles,
and open triangles show the results for $M_1/M_2=1, t_{e,1}/T_{\rm K}=10^3,
t_{e,2}/T_{\rm K}=\infty,$ and $e_{\rm ini}=10^{-2},$ respectively, but the other parameters
are the same as for the fiducial case. Note that the open squares and open
circles overlap at every $M_1$. 
The dependence on $M_{1}$ is examined between $M_{1}/M_\oplus=10^{-1}$ and  $10^{2}$.
In order to keep computational cost reasonable, we also put an upper limit of 
$t_{a,{\rm crit}}=10^{8} T_{\rm K}$. For $e_{\rm ini}=10^{-2}$, the eccentricity of the
inner body is $\simeq 0.01$ at the resonant encounter and the capture probability curve is broadened
($\gamma \simeq -2$--$-4$).

This survey reveals two features. First, all of the $t_{a,{\rm crit}}$
values except that for $M_1/M_2=1$
are within about a factor of two of each other. This suggests that 
$t_{a,{\rm crit}}$ 
depends weakly on $t_{e,1}, t_{e,2},$ and $e_{\rm ini}$, which can be seen in the
following subsections in more detail. 
Second, all the cases exhibit a similar power--law dependence 
of $t_{a,{\rm crit}}$ on $M_1$. The gradients
are calculated for each case by least--squares fits, and we obtain a typical power
index approximately equal to $-4/3$.
A physical interpretation for this dependence is discussed
in Section~\ref{sec:discussion}.

\subsection{Dependence on $M_2/M_1$}
\begin{figure}[htbp]
\epsscale{0.9}
\plotone{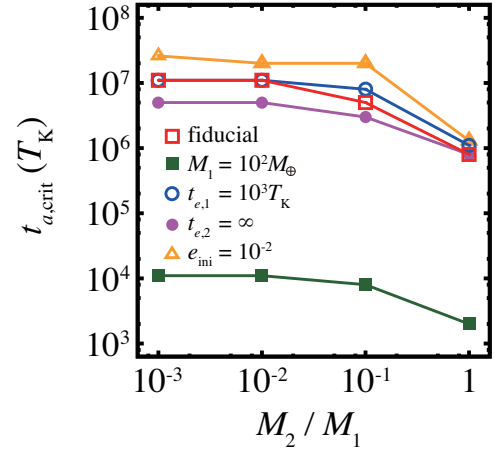}
\caption{Critical migration timescale versus $M_2/M_1$. Open squares represent the fiducial model.
Filled squares, open circles, filled circles,
and open triangles represent the results for $M_1/M_\oplus=10^2, t_{e,1}/T_{\rm K}=10^3, 
t_{e,2}/T_{\rm K}=\infty$  and $e_{\rm ini}=10^{-2},$ respectively.
}
\label{fig:m2_ta}
\end{figure}

Next, we explore the effect of varying $M_2/M_1.$ Note that the dependence
can be seen more clearly when $M_2/M_1$ is used instead of $M_2$.
Figure~\ref{fig:m2_ta} shows $t_{a,{\rm crit}}$ as a function of $M_2/M_1.$
Open squares represent the fiducial case. Filled squares, open circles, filled circles,
and open triangles show the results for $M_1/M_\oplus=10^2, t_{e,1}/T_{\rm K}=10^3, 
t_{e,2}/T_{\rm K}=\infty$, and $e_{\rm ini}=10^{-2},$ respectively. 
Same as the previous subsection, the capture probability curve is not 
sharp for $e_{\rm ini}=10^{-2}$ ($\gamma \simeq -1$--$-3$).

We find that if $M_2/M_1 \lesssim 10^{-1},$ 
$t_{a,{\rm crit}}$ is independent of $M_2/M_1$; therefore, the restricted three--body
approximation is valid for this condition.  
The difference between $t_{a,{\rm crit}}$ for $M_2/M_1=1$ and the value for small
$M_2/M_1$ is found to be a factor of about 10. Thus, although the
restricted three--body approach cannot provide an accurate prediction of $t_{a,{\rm crit}}$
for bodies with comparable masses, it would be possible to roughly
derive $t_{a,{\rm crit}}$ to an accuracy of a factor of 10.
See also discussions in Sections~\ref{sec:internal_resonance} and
\ref{sec:compare} for the reason for the difference.

Note that we consider only the migration of the outer
planet. Because migration speed depends on planetary mass, the inner planet's migration is not negligible,
especially for $M_1 \sim M_2$.  
However, if we apply the differential migration speed instead of the
migration speed of the outer planet, the critical migration timescales
we obtain are valid even considering the migration of both planets.

\subsection{Dependence on $t_e$}
\begin{figure}[htbp]
\epsscale{0.9}
\plotone{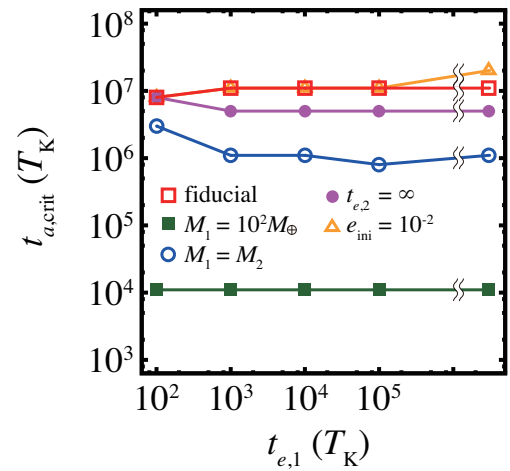}
\caption{Critical migration timescale versus $t_{e,1}$. Open squares are the fiducial case.
Filled squares, open circles, filled circles,
and open triangles represent the results for $M_1/M_\oplus=10^2, M_1/M_2=1,
t_{e,2}/T_{\rm K}=\infty$,  and $e_{\rm ini}=10^{-2},$ respectively.
}
\label{fig:te1_ta}
\end{figure}

Figure~\ref{fig:te1_ta} shows the results for various eccentricity--damping timescales
for the inner (larger) planet, $t_{e,1}$. Again, the open squares represent the
fiducial case. Filled squares, open circles, filled circles,
and open triangles show the results for $M_1/M_\oplus=10^2, M_1/M_2=1,
t_{e,2}/T_{\rm K}=\infty$,  and $e_{\rm ini}=10^{-2},$ respectively.
The rightmost points on the horizontal axis are the cases without eccentricity damping.
For $e_{\rm ini}=10^{-2}$ and $t_{e,1}=\infty$, the transition from a capture 
probability of zero to one is not very sharp ($\gamma \simeq -2$).
There is no systematic change in $t_{a,{\rm crit}}$ with $t_{e,1}.$
Even when $t_{e,1}$ is varied
by more than four orders of magnitude, the differences in $t_{a,{\rm crit}}$ lie within
a factor of two or three.

\begin{figure}[htbp]
\epsscale{0.9}
\plotone{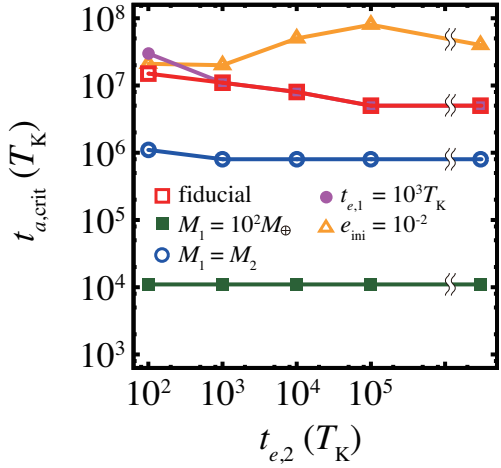}
\caption{Critical migration timescale versus $t_{e,2}$. Open squares represent the fiducial case.
Filled squares, open circles, filled circles,
and open triangles represent the results for $M_1/M_\oplus=10^2, M_1/M_2=1,
t_{e,1}/T_{\rm K}=10^3$,  and $e_{\rm ini}=10^{-2},$ respectively.
}
\label{fig:te2_ta}
\end{figure}

Figure~\ref{fig:te2_ta} shows the results for various eccentricity--damping timescales
for the outer (smaller) planet, $t_{e,2}$. Open squares represent the fiducial case.
Filled squares, open circles, filled circles,
and open triangles are the results for $M_1/M_\oplus=10^2, M_1/M_2=1,
t_{e,1}/T_{\rm K}=10^3$, and $e_{\rm ini}=10^{-2},$ respectively.
For $e_{\rm ini}=10^{-2}$, the transition in the capture protability is not sharp
($\gamma \simeq -2$--$-6$).
We also do not see any clear trends in $t_{e,2}.$
Although the case with $e=10^{-2}$ (open triangles) shows a slight variation, it still remains
within a factor of a few.

\subsection{Dependence on $e_{\rm ini}$}
\label{sec:eini}
\begin{figure}[htbp]
\epsscale{0.9}
\plotone{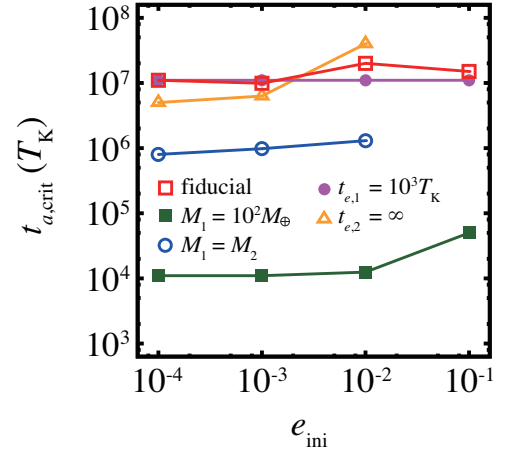}
\caption{Critical migration timescale versus initial eccentricity. Open squares represent the fiducial case.
Filled squares, open circles, filled circles,
and open triangles represent the results for $M_1/M_\oplus=10^2, M_1/M_2=1,
t_{e,1}/T_{\rm K}=10^3$  and $t_{e,2}/T_{\rm K}=\infty,$ respectively.
}
\label{fig:e_ta}
\end{figure}

Finally, Figure~\ref{fig:e_ta} shows the results for various initial eccentricities
$e_{\rm ini}$. Open squares represent the fiducial case.
Filled squares, open circles, filled circles,
and open triangles are the results for $M_1/M_\oplus=10^2, M_1/M_2=1,
t_{e,1}/T_{\rm K}=10^3$,  and $t_{e,2}/T_{\rm K}=\infty,$ respectively.
When the eccentricity is not small $(e \ga 0.01)$ at the resonant encounter,
in other words, when $e_{\rm ini} \geq 10^{-2}$ and $t_{e} > t_{a}$,
the capture probability curve is not sharp the same as before.
Again, no clear trends in $t_{a,{\rm crit}}$ with $e_{\rm ini}$ are recognized.
Note that, for $e_{\rm ini}=10^{-1}$ and $M_1/M_2=1$ (open circles) 
and for $t_{e,2}/T_{\rm K}=\infty$ (open triangles), $t_{a,{\rm crit}}$ cannot 
be determined to have particular values. Because the eccentricity of the smaller body $M_2$
at a resonant encounter is relatively large, adequate resonant capture does not occur 
even for long $t_a$;
the capture probability does not exceed about 0.3.
Except for large $e_{\rm ini}$, the dependence is reasonably weak.
Several of these features are in agreement with previous studies, which will be
discussed in Section~\ref{sec:compare}.

\subsection{Summary of Results}
The critical migration timescales derived in the previous subsections are summarized as
follows:
\begin{eqnarray}
t_{a,{\rm crit}} \simeq \left\{ \begin{array}{ll}
1\times 10^7 \left(\frac{M_1}{M_\oplus}\right)^{-4/3}\left(\frac{M_*}{M_\odot}\right)^{4/3}
~T_{\rm K}& (M_2/M_1 \lesssim0.1)\\
1\times 10^6 \left(\frac{M_1}{M_\oplus}\right)^{-4/3}\left(\frac{M_*}{M_\odot}\right)^{4/3}
~T_{\rm K}& (M_2/M_1 \simeq 1),\\
\end{array} \right.
\label{eq:ta_crit}
\end{eqnarray}
where the dependence on the stellar mass $M_*$ is included.
As shown in the results, $t_{a,{\rm crit}}$ can differ by a factor of two or three.
Although we assume that the inner body is more massive than the outer body
$(M_1 \geq M_2)$, the results  are almost the same when the outer one
is more massive, which will be seen in Section~\ref{sec:internal_resonance}.
The formula is valid if the eccentricities of the smaller planets at a resonant
encounter are much smaller than 0.1.

\subsection{Closely Spaced Resonances}
\label{sec:closer_resonances}
The capture condition for 2:1 mean motion resonances was examined above. 
We performed additional simulations to evaluate the critical
migration timescale for closely spaced first--order resonances (e.g., 3:2 and 4:3) because several
exoplanet systems have such commensurabilities.

\begin{figure}[htbp]
\epsscale{0.9}
\plotone{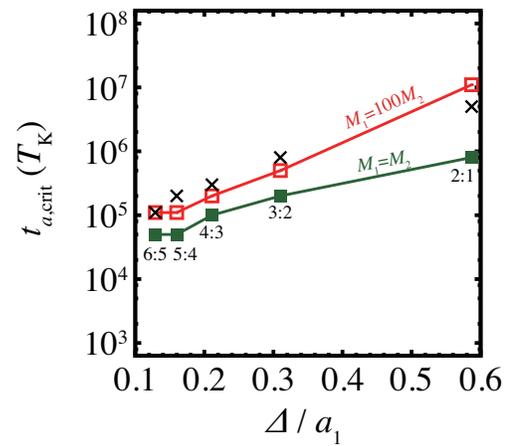}
\caption{Critical migration timescales for 2:1 and more closely spaced resonances.
Open and filled squares represent the results for $M_2/M_1 = 10^{-2}$ and
$M_2/M_1 = 1$, respectively. Crosses represent the results for internal
resonances ($M_1/M_\oplus = 10^{-2}$ and $M_2/M_\oplus = 1$).}
\label{fig:p_ta}
\end{figure}

Figure~\ref{fig:p_ta} shows the critical migration timescales for $p+1:p$ resonances ($p=1, 2, 3, 4, 5$)
obtained by numerical simulations. Here, we define 
\begin{equation}
\Delta \equiv \left[\left(\frac{p+1}{p}\right)^{2/3} - 1
		     \right]a_1, 
\end{equation}
which roughly corresponds to the orbital separation from the inner
planet to the outer planet in the mean motion resonance. 
Open squares represent
the results for $M_2/M_1=10^{-2}$, whereas filled squares represent those for equal--mass
bodies ($M_2/M_1=1$). In each case, the other parameters are set to the fiducial values
$(M_1/M_\oplus=1, t_{e,1}/T_{\rm K}=\infty, t_{e,2}/T_{\rm K}=10^3, e_{\rm ini}=10^{-4})$.
We see that the critical migration time decreases with decreasing separation.
The critical migration timescale for equal--mass bodies ($M_1=M_2$) is always shorter than that
with bodies that have a high mass ratio ($M_1 = 100 M_2$).
The difference in $t_{a,{\rm crit}}$ is larger at 2:1 resonances than 
at closely spaced resonances. This is presumably because the strength of the 2:1 external 
resonance, where ``external'' means that the inner body is the dominant body, is significantly 
weakened by indirect terms in the disturbing function, which will also be discussed
in Section~\ref{sec:internal_resonance}.

\begin{figure}[htbp]
\epsscale{1}
\plotone{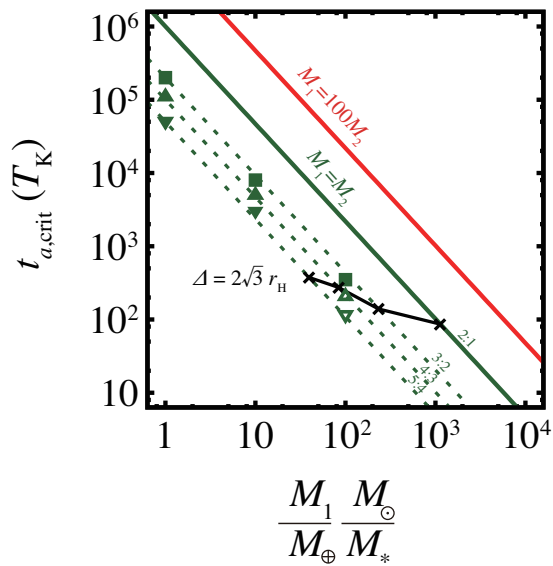}
\caption{Summary of critical migration timescale derived from numerical simulations.
Solid and dashed lines represent the critical migration timescales for 2:1 and
more closely spaced resonances, respectively. Filled symbols indicate the numerically
determined values, whereas for open symbols, $t_{a,{\rm crit}}$ cannot be derived owing to
close encounters. The orbital separation $\Delta$ divided by the mutual Hill radius 
$r_{\rm H}$ can be expressed in terms of the mass and semimajor axis. The masses of
$\Delta = 2\sqrt{3}~r_{\rm H}$ for each resonance are plotted with crosses and connected 
by a solid line by assuming $M_1 =  M_2$.}
\label{fig:dis2_m1_ta}
\end{figure}

\begin{deluxetable}{cccccc}
\tablecolumns{8}
\tablewidth{0pc}
\tablecaption{Fitting results for $C$}
\startdata
\hline \hline
Mass ratio	& 2:1	& 3:2	& 4:3	& 5:4	&6:5\\
\hline
$M_2/M_1 \simeq 1$	& $1 \times 10^6$	& $2 \times 10^5$	& $1 \times 10^5$	& $5 \times 10^4$	& $5 \times 10^4$\\
$M_2/M_1 \lesssim 0.1$	& $1 \times 10^7$	& $5 \times 10^5$	& $2 \times 10^5$	& $1 \times 10^5$	& $1 \times 10^5$
\enddata
\tablecomments{
Fitting results to Equation~(\ref{eq:ta_gen}) for each mean motion resonance.
Note that, for $M_2/M_1 \la 0.1$ and the 3:2, 4:3, 5:4 and 6:5 resonances,
the mass dependence
is not examined, but the critical migration timescales are 
derived only for $M_1 = 1 M_\oplus$ and $M_2 = 10^{-2} M_\oplus$.}
\label{tbl:fitting}
\end{deluxetable}

The combined results from the above surveys are plotted in Figure~\ref{fig:dis2_m1_ta}.
The solid lines indicate the critical migration timescale for capture into the 
2:1 resonance as a function of $M_1$, which is summarized in Equation~(\ref{eq:ta_crit}).
Squares, triangles, and inverted triangles are the numerical results for the
3:2, 4:3, and 5:4 resonances, respectively.
From the results, the critical migration timescale is described as 
\begin{equation}
 t_{a,{\rm crit}} = C \left(\frac{M_1}{M_\oplus}\right)^{-4/3}
  \left(\frac{M_*}{M_\odot}\right)^{4/3} T_{\rm K},\label{eq:ta_gen} 
\end{equation}
where $C$ depends on $M_1/M_2$ and the resonant commensurability. 
We derive the $C$ values from the fitting of the results and 
summarize $C$ in Table~\ref{tbl:fitting}. 
Note that although $C$ should depend on commensurability,
the values are the same between the 5:4 and 6:5 resonances, which suggests that
the difference is within the minimum interval of $t_a$ that we set for our investigation.
The dashed lines in Figure~\ref{fig:dis2_m1_ta} represent fitting results for
equal--mass cases $(M_1 = M_2)$.
Although the approximate fits for closer resonances are considered to be correct
within a factor of a few, they should be treated with some caution.
We observe that when $M_1/M_\oplus=100$ and $p=3, 4$, the bodies undergo 
close encounters before being captured into resonances; therefore, 
the bodies do not settle into stable resonant orbits, as indicated by the
open symbols in Figure~\ref{fig:dis2_m1_ta}.

According to the analysis of the Hill stability by \citet{gladman93}, dynamical
stability is almost guaranteed if $\Delta \gtrsim 2 \sqrt{3}~r_{\rm H}$, where $r_{\rm H}$ 
is the mutual Hill radius. 
Note that if the orbital separation is within a few tens of percent beyond the 
critical Hill separation, Hill--stable planetary systems may manifest Lagrange instability when
the outer planet escapes to infinity (\citealt{barnes_greenberg05}; \citealt{veras_mustill13}).
There is no analytical criteria that describe Lagrange stability, and additional numerical 
simulation is required to determine whether the system is Lagrange stable or not. However, 
\citet{barnes_greenberg05} found that the Hill stable condition is a good predictor of Lagrange
stability.
The Hill stable condition is rewritten as 
\begin{equation}
M_1 + M_2 \leq  
\frac{1}{\sqrt{3}} 
\left[\frac{\left(p+1\right)^{2/3} - p^{2/3}}
{\left(p+1\right)^{2/3} + p^{2/3}}\right]^3
M_*.
\label{eq:critical_mass}
\end{equation}
The critical masses for equal--mass bodies assuming $M_* = M_\odot$
are plotted with crosses connected with by a solid line in Figure~\ref{fig:dis2_m1_ta}.
If the mass of the body is larger than the critical mass, the system becomes
Hill unstable, which is consistent with our results that exhibit close encounters
(open symbols).
Note that even if $\Delta \la 2 \sqrt{3} r_{\rm H}$, the orbit in a resonance 
can become stable for specific orbital arguments (some examples are provided in Section~\ref{sec:constraints}).
This criterion for the instability is a rough estimate, and 
long--term orbital integration can reveal the instability time
of planets in closely spaced mean motion resonances.

\subsection{Internal Resonances}
\label{sec:internal_resonance}
So far, we have carried out simulations in the cases where the inner body is more massive or 
equal to the outer body.
In order to evaluate the difference in $t_{a,{\rm crit}}$ between the internal and external
resonances, additional simulations for outer massive body are performed.
Assuming $M_1=10^{-2} M_\oplus$, $M_2=1 M_\oplus$, 
$t_{e,1}=10^{3} T_{\rm K}$, $t_{e,2}=\infty$, and $e_{\rm ini}=10^{-4}$,
the critical migration timescales are determined for 2:1, 3:2, 4:3, 5:4, and 6:5 resonances,
which are plotted in Figure~\ref{fig:p_ta} with crosses.

We find that no significant difference between the internal and external resonances; in fact,
it lies within a factor of two. Therefore, 
as stated in the last sentence of Section~\ref{sec:model},
if the outer body is larger than the inner body, 
we apply the mass of the larger body to $M_1$; Equation~(\ref{eq:ta_gen}) is then valid.

Although the difference is not significant, we see a decrease in $t_{a,{\rm crit}}$
at the 2:1 internal resonance, which means that the 2:1 internal resonance is stronger
than the 2:1 external resonance.
This tendency can be understood in terms of contributions of
each term in the disturbing function. For the 2:1 external resonance, the contribution of
the direct term is diminished by the indirect term, leading to weakening the strength of the
resonance (\citealt{quillen06}; \citealt{mustill_wyatt11}). One indicator of the strength of
the resonance is $l_j$ in the work of \citet{mustill_wyatt11} (see Table~1 in their work
for each value), where the strength of resonance decreases with increasing $l_j$.
We see that $l_j$ for the 2:1 external resonance is large. The fact that the 2:1 internal
resonance is stronger than the 2:1 external resonance would also partially explain 
the decrease in
$t_{a,_{\rm crit}}$ for equal--mass bodies in Figures~\ref{fig:m2_ta} and \ref{fig:p_ta}
although this cannot account for the entire change.
We also see in Figure~\ref{fig:p_ta} that internal resonances are slightly
weaker than external resonances for closely spaced resonances (e.g., 3:2), which is 
consistent with Figure~11 in the work of \citet{mustill_wyatt11}.

\section{ANALYSIS OF THE DEPENDENCE OF $t_{a,{\rm crit}}$ ON $M_1$}
\label{sec:discussion}

We discuss the numerical results using analytical arguments.
Through numerical investigations, we find that the critical migration timescale shows
a power--law behavior, $t_{a,{\rm crit}} \propto (M_1/M_*)^{-4/3}$, which was also seen
in previous studies (e.g., \citealt{ida_etal00}; \citealt{quillen06}). This tendency can be
estimated using a simple pendulum model \citep{murray_dermott99}; the following
expressions describe the orbital properties of a massless particle that is in a mean motion
resonance. Note that 
these expressions are for the internal resonance; however, those for the external resonance
are almost the same.
For the circular restricted problem, where the massless test particle is in a
first--order $p+1:p$ resonance with a body with mass of $M$, the maximum width of
libration is given by
\begin{eqnarray}
\Delta_{\rm res} = \left[ \frac{-16 C_{\rm r} e_{\rm res}}{3 n}\right]^{1/2} a,
\label{eq:delta_res}
\end{eqnarray}
where $e_{\rm res}$ is the excited eccentricity due to resonant
perturbation, and $n$ and $a$ are the mean motion  and semimajor axis of the test particle,
respectively. The 
constant arising from the resonant term of the disturbing function is
\begin{eqnarray}
C_{\rm r} 
&=& -\frac{M}{M_*} n \alpha \left(p+1 + \frac{\alpha}{2} D\right)b_{1/2}^{p+1},\\
&=& \frac{M}{M_*} n \alpha f_{\rm d}(\alpha),
\end{eqnarray}
where $\alpha, b_{1/2}^{p+1}$, and $D$ are  
the ratio of semimajor axis of the inner body to that of the outer body,  
the Laplace coefficient, and the derivative operator, respectively. For the
first--order resonance, $C_{\rm r}$ is always negative.
The libration timescale is derived as
\begin{eqnarray}
\tau_{\rm lib} = \frac{2 \pi}{(-3 p^2 C_{\rm r} n e_{\rm res})^{1/2}}.
\label{eq:t_lib}
\end{eqnarray}
Then the excited eccentricity during the resonant passage is approximated
using an adiabatic invariant as 
\begin{eqnarray}
e_{\rm res}^2 = \frac{\Delta_{\rm res}}{p a},
\end{eqnarray}
as in \citealt{zhou_lin07}.
Substituting this equation into Equations~(\ref{eq:delta_res}) and (\ref{eq:t_lib}),
we obtain
\begin{eqnarray}
e_{\rm res} = \left( \frac{-16 C_{\rm r}}{3 p^2 n}\right)^{1/3},
\end{eqnarray}
\begin{eqnarray}
\Delta_{\rm res} = \left( \frac{-16 C_{\rm r}}{3 n}\right)^{2/3} \frac{a}{p^{1/3}},
\end{eqnarray}
\begin{eqnarray}
\tau_{\rm lib} = \left( \frac{n^2}{12 p^2 C_{\rm r}^2}\right)^{1/3} T_{\rm K}.
\end{eqnarray}
Comparing the libration timescale $\tau_{\rm lib}$ and the migration timescale through the 
resonant width $\tau_a$, where
\begin{eqnarray}
\tau_a = \frac{\Delta_{\rm res}}{\dot{a}} = \frac{\Delta_{\rm res} t_a}{a},
\end{eqnarray}
we determine that the critical migration timescale can be roughly given by
\begin{eqnarray}
t_{a,{\rm crit}} = \left( \frac{- 3}{1024 p \alpha^4 f_{\rm d}(\alpha)^4}\right)^{1/3}
\left(\frac{M}{M_*}\right)^{-4/3} T_{\rm K}.
\end{eqnarray}
We confirm the power--law dependence on the mass. In addition, given that
$p=1$, $\alpha = 0.63$, and $\alpha f_{\rm d}(\alpha) \simeq -0.75$ \citep{murray_dermott99},
$t_{a,{\rm crit}}$ for an Earth--mass body is $\sim 10^7 T_{\rm K}$.
These estimates are roughly consistent with our numerical results; note, however, 
that they are only  order--of--magnitude estimates assuming the restricted problem.
The contribution from the variation in the longitude of the pericenter $\varpi$ is also neglected.

\section{COMPARISON WITH PREVIOUS STUDIES}
\label{sec:compare}
We here discuss our results by comparing with previous studies. First, our results for
two bodies that have a high mass ratio
can be compared to the works that consider the restricted problem
(e.g., \citealt{quillen06}; \citealt{mustill_wyatt11}). We find that several features in 
Figure~\ref{fig:e_ta}, in which the dependence of $t_{a,{\rm crit}}$ on the eccentricity is
investigated, are in agreement with the results of \citet{mustill_wyatt11} (Figure~2 in their
paper) as follows. \citet{mustill_wyatt11} showed that when the generalized momentum 
is small enough, which corresponds to small $e$, the capture probability curve
is steep and $t_{a,{\rm crit}}$ hardly depends on $e$. In Figure~\ref{fig:e_ta}, we observe
that $t_{a,{\rm crit}}$ is well--defined and almost independent of $e$ when $e$ at the 
resonant encounter is small ($e \la 10^{-3}$ for $M_1 = 1 M_\oplus$).
In this case, the value of $t_{a,{\rm crit}}$ is also comparable to that derived by \citet{mustill_wyatt11};
that is, their estimate for $t_{a,{\rm crit}}$ is $\simeq 4 \times 10^6 T_{\rm K}$ for 
$M_1 = 1 M_\oplus$. 
We also see another feature of steepening dependence on $e$ at higher $e$. According
to \citet{mustill_wyatt11}, $t_{a,{\rm crit}}$ tends to sensitively depend on $e$ when
$e$ is larger than $\sim 0.01$ and $\sim 0.05$ for $M_1 = 1 M_\oplus$ 
and $100 M_\oplus$, respectively, which are seen in Figure~\ref{fig:e_ta} in the models of
$t_{e,2}=\infty$ and $M_1 = 10^2 M_\oplus$. The other feature is that the capture
probability has a cutoff at high eccentricity and never exceeds
about 0.3 in the case of $t_{e,2}=\infty$. In our results, the cutoff eccentricity is 
$e \simeq 0.01$, which quantitatively reproduces the results of \citet{mustill_wyatt11}.

In addition, several similar features are observed even in the results for equal--mass bodies;
namely, the weak dependence of $t_{a,{\rm crit}}$ on $e$ and the existence of the cutoff 
in the capture probability at high $e$, which were expected from the previous study
\citep{mustill_wyatt11}. Note, however, that the values of $t_{a,{\rm crit}}$ differ by a factor
of 10 between the restricted problem and the unrestricted problem. Even if we consider
the difference between the internal resonance and the unrestricted problem, there is 
at least a factor of a few difference in $t_{a,{\rm crit}}$.
Special care would have to be taken when one quantitatively discuss the critical migration 
timescale for equal--mass bodies.

Several properties observed in our simulations were also seen in a dynamical study by \citet{rein_etal12}.
Figure~2 of \citet{rein_etal12} indicates the excluded resonant regions for the HD~200964 system;
Figure~2(a) shows their results for two planets of comparable mass, whereas Figure~2(c)
shows the case where one body is a test particle. They found that the allowed
region for the comparable--mass case is wider than that for the zero--mass particle,
which suggests that planets with comparable masses are more easily captured into 
mean motion resonances.
In our model, we find that the critical migration timescale for equal--mass planets is shorter 
than that for systems with high mass ratios.
In addition, we also see a weak dependence of the critical migration timescale on
the $e$--damping timescales. This tendency is also seen in Figure~4 of \citet{rein_etal12}.

\citet{papaloizou_szuszkiewicz05} derived an analytical expression for the eccentricity
of the outer planet captured into a resonance.
The equilibrium eccentricity is attained by balancing pumping due to
resonant effects with damping caused by migration, as described by Equation~(5) of 
\citet{papaloizou_szuszkiewicz05}.
We confirmed that our numerical results are consistent with this analytical formula except 
for rapid migration $(t_e/t_a \gtrsim 0.1)$ and equal--mass bodies.

\section{CONSTRAINTS ON FORMATION MODELS}
\label{sec:constraints}
The results given in Section~\ref{sec:results} can be used to constrain the history of
systems in mean motion resonances and their formation scenarios.
In this section, we first discuss exoplanet systems with resonances 
using the mass versus critical migration timescale 
diagram, and then move on to other systems.

\subsection{Exoplanet Systems}
As of 2013, more than 25 planetary systems that lie in or close to first--order mean motion resonances
have been detected (e.g., \citealt{steffen_etal13}). 
Table~\ref{tbl:exoplanets} lists the orbital properties of the confirmed planets that could be in 
first--order resonances\footnote{This is taken from two papers 
(\citealt{szuszkiewicz_podlewska-gaca12}; \citealt{steffen_etal13}) and should not be a complete list.}.
The fourth and fifth columns show the masses and semimajor axes of the planets, respectively;
the sixth and seventh columns show the resonant properties. For example, the third row indicates
that Gliese~876~c could be in a 2:1 resonance with planet b, where the period ratio of
these planets is 2.03. The data are taken from the Open Exoplanet 
Catalogue\footnote{http://exoplanet.hanno-rein.de/}.
Note that many systems listed have ratios more distant from
commensurability than the 1\% used to classify resonant captures in our simulations.
Although the period ratios are close to commensurate values, not all of the planets 
necessarily lie in mean motion resonances.

\begin{figure}[htbp]
\epsscale{1.1}
\plotone{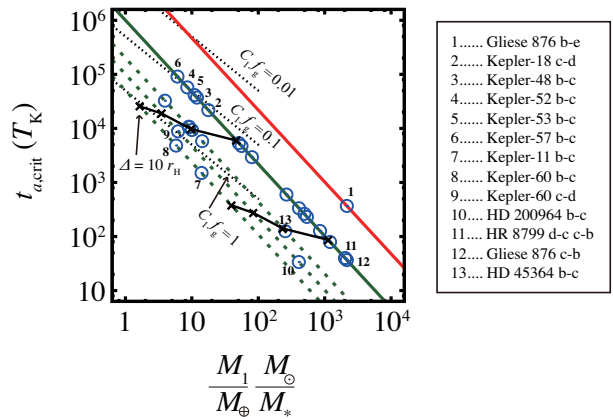}
\caption{Same as Figure~\ref{fig:dis2_m1_ta}, but observed resonant pairs listed in
Table~\ref{tbl:exoplanets} are also plotted as open circles, which indicate the critical
migration timescale for each pair. This means that the pair would be moved in the gas
disk on a relative migration timescale longer than the critical value.
In addition to $\Delta = 2 \sqrt{3} r_{\rm H}$, $\Delta = 10 r_{\rm H}$ is also marked
by crosses connected by a solid line, which indicate the typical orbital separation
of planets formed via oligarchic growth.
The three dotted lines represent the type I migration timescale with different migration
efficiencies, $C_{\rm I}f_{\rm g} = 1, 0.1,$ and 0.01.}
\label{fig:dis_m1_ta}
\end{figure}

The solid and dashed lines in Figure~\ref{fig:dis_m1_ta}, which indicate the 
critical migration speeds,
are the same as in Figure~\ref{fig:dis2_m1_ta}.
For the resonant--pair exoplanets listed in Table~\ref{tbl:exoplanets},
the critical migration timescale, plotted as open circles, is obtained from
their resonant commensurability and the mass ratio between the smaller
and larger bodies ($M_2/M_1 <0.1$ or not) as a function of the larger planet mass. 
All pairs except for the pair of Gliese~876 b and e have roughly equal masses
$M_2/M_1 > 0.1$.
The outer planet migration timescale must be longer than the critical migration
time for capture into mean motion resonance to occur. 
We obtained the critical migration timescale in Section~\ref{sec:results} while 
ignoring the migration of inner planets. If the inner planets also migrate inward, 
the differential (or relative) migration speed between the inner and outer planets should be applied to the critical migration speed. 

Using Figure~\ref{fig:dis_m1_ta}, we can place some constraints on the differential migration 
speeds that the exoplanets used to have. 
Note that in the following discussions, we introduce several assumptions: 
(i) Planets formed widely separated from the 2:1 resonance. (ii) The masses of planets do 
not evolve after the onset of migration. (iii) Migration is smooth and not subjected to 
stochastic
torques\footnote{\citet{rein12} have shown that orbital integrations that include 
some degree of stochasticity is necessary to reproduce the orbital period ratio distribution of the 
\textit{Kepler} planet candidates. We assume smooth migration in this paper
for simplicity.}. (iv) The eccentricities of planets are low and hence the transition from
certain capture to certain failure is well--defined. (v) Once captured in resonance,
planets do not subsequently escape from the resonance.

The 2:1 mean motion resonance is the outermost 
first--order mean motion resonance. The capture probability for
the outer, higher--order mean motion resonance is very low. 
Therefore, if a system has a 2:1 commensurability, the differential migration speed would have been 
slower than that estimated from the critical migration time. 
If a system has a 3:2 commensurability, the outer planet passed through the 2:1 mean motion 
resonance because of the high migration speed and was captured into the 3:2 mean motion resonance; 
thus, the differential migration speeds is slower than the critical migration speed for the 3:2 mean 
motion resonance but faster than that for the 2:1 mean motion resonance. 
We can constrain the differential migration speed for capture
in the other mean motion resonances in a similar manner.  The pairs of exoplanets in
the closely spaced mean motion resonances are expected to have migrated at significantly high speed. 

Note that if two planets in a mean motion resonance formed with a small orbital 
separation, they may have been captured into closely spaced resonances under slow migration. 
In addition, if several planets 
simultaneously exhibit migration as a resonant convoy (\citealt{mcneil_etal05};
\citealt{ogihara_ida09}) rather than the single outer planet migrating inward, the critical migration timescale
can be somewhat longer than that derived in Section~\ref{sec:results}.
Nonetheless, our approach provides useful constraints on formation models 
without the need for further dynamical analyses and calculations for individual systems.

\subsubsection{Overall Trend}
Although the number of observed resonant exoplanets is too small to support a statistical discussion,
we see in Figure~\ref{fig:dis_m1_ta} that there is a general trend toward a decrease in the
number of systems in closely spaced resonances with increasing mass.
This trend can be understood in terms of the short critical migration timescale for 
capture into 2:1 resonances with high--mass planets; 1000--Earth--mass planets 
can capture bodies into 2:1 resonances even if the migration timescale is
quite short $(\simeq 100 T_{\rm K})$.

As stated in Section~\ref{sec:closer_resonances}, it is predicted that pairs with
orbital separations smaller than $\simeq 2 \sqrt{3}$ Hill radii can be Hill unstable, and 
such configurations are rare. We note that the Hill stability criterion, more properly
bodies should be Lagrange stable, gives only
a sufficient condition for stability; thus, 
if the planets are in mean motion resonances, the system can become
stable. However, they should experience a configuration where planets are 
not near resonances at some point during migration phase; therefore, 
it is possible that such planets undergo
close encounters and are scattered away from the system.
We find that almost all the pairs have separations larger than 
$\simeq 2 \sqrt{3} r_{\rm H}$; however, there are several exceptions. We discuss these systems 
in Section~\ref{sec:small_separation}.

On the other hand, several pairs have separations larger than $\simeq 10$ Hill radii,
which is the typical orbital separation of planets formed via the oligarchic growth phase
\citep{kokubo_ida98}.
The location of $\Delta = 10 r_{\rm H}$ is also shown in Figure~\ref{fig:dis_m1_ta} 
by crosses connected by a solid line, in the same way as for $\Delta = 2 \sqrt{3}~r_{\rm H}$.
Thus, planets are believed to
form in distant orbits, after which they migrate inward one by one, maintaining their orbital
separations wider than $10 r_{\rm H}$ \citep{ogihara_ida09}.
These systems include Kepler-18, Kepler-48, Kepler-52, Kepler-53, and Kepler-57,
which all have 2:1 commensurabilities. Figure~\ref{fig:dis_m1_ta} gives the
lower limits on the migration timescale.

For example, the Kepler-18 system consists of two low--density Neptune--mass planets (c and d) near a 
2:1 resonance and an inner super--Earth (b) \citep{cochran_etal11}. The orbital separation
of planets c and d is $\simeq 13 r_{\rm H}$. The planets are believed to undergo
migration with a differential migration speed slower than $2 \times 10^4 T_{\rm K}$.

Next, we focus on super--Earth--mass planets $(M \sim 10 M_\oplus)$, which have a relatively
large number of samples, and compare their histories with the migration theory.
Planets with a few tens of the Earth mass or less might have experienced type I migration.
Through a linear calculation, the type I migration timescale is given by 
\citep{tanaka_etal02}
\begin{eqnarray}
t_{a,{\rm lin}} = 
\frac{1}{2.7 + 1.1q} \left(\frac{M}{M_*}\right)^{-1}
\left(\frac{\Sigma_{\rm g} r^2}{M_*}\right)^{-1}
\left(\frac{c_{\rm s}}{v_{\rm K}}\right)^2~\Omega_{\rm K},\label{eq:type1lin}
\end{eqnarray}
where $-q$ denotes the surface density gradient. 
For optically thin disks,
the temperature distribution is \citep{hayashi81}
\begin{equation}
T \simeq 280 \left(\frac{r}{1 {\rm AU}}\right)^{-1/2}
\left(\frac{L_*}{L_\odot}\right)^{1/4}~{\rm K},
\end{equation}
which determines the sound velocity $c_{\rm s}$. 
We scale the gas surface density
as
\begin{equation}
\Sigma_{\rm g} = 2400 f_{\rm g}  \left(\frac{r}{1 {\rm AU}}\right)^{-q}~{\rm g~cm^{-2}},
\end{equation}
where $f_{\rm g}$ is a scaling factor. 
If we introduce the type I migration efficiency $C_{\rm I} \equiv t_{a,{\rm
lin}}/t_a$ to express the uncertainty in the type I migration theory,
the type I migration timescale for $q=3/2$ and $L_* = L_\odot$ is
written as 
\begin{equation}
\label{eq:typeI}
t_a = 5.0 \times 10^4 C_{\rm I}^{-1} f_{\rm g}^{-1} \left(\frac{M}{M_\oplus}\right)^{-1}
T_{\rm K}.
\end{equation}
The type I migration speed is still uncertain\footnote{In an optically thick
disk, the migration timescale can be long, and outward migration is
possible under some conditions \citep{paardekooper_etal11}. 
A population synthesis model by \citet{ida_lin08} suggests that the typical migration
efficiency is less than that obtained by linear analysis ($C_{\rm I} \la 0.1$).}.
Here, we consider a scenario in which the inner body is stationary and the outer body
undergoes inward migration. Then the type I migration timescale (Equation~[\ref{eq:typeI}]),
which is identical to the relative migration timescale between the two bodies,
can be drawn as a function of $C_{\rm I}f_{\rm g}$ (dotted line in Figure~\ref{fig:dis_m1_ta}).

Some of the exoplanets with masses of $\sim 10 M_\oplus$
are in 2:1 resonances. As shown in
Figure~\ref{fig:dis_m1_ta}, they would not have undergone
rapid migration, so $C_{\rm I} f_{\rm g} \la 0.1$. On the other hand, some planets
are captured into closely spaced resonances. 
For them, a short migration time is required, and $C_{\rm I} f_{\rm g} \sim 1$. 

In order to reconcile the low migration efficiency ($C_{\rm I} f_{\rm g} \la 0.1$)
and the linear type I migration theory, the migration
of inner planets should be considered. If the inner planets migrate when they
are captured into mean motion resonances, the differential migration speeds become much
slower than the migration rate estimated from Equation~(\ref{eq:type1lin}). 
This can explain the planets in the 2:1 mean motion resonances. 
On the other hand, the migration of inner planets is negligible 
if the inner planets are much smaller than the outer ones and/or 
if the inner planets have orbits around the inner edge of the disk. 
Such pairs of planets can be in closely spaced resonance because of the
high relative migration speed. 

In addition, the other possibility for a pileup of planets with masses of $\simeq 10 M_\oplus$ is 
that they are formed near the current
resonance locations (e.g., 3:2 and 2:1). The orbital separations for 10--Earth--mass planets in 3:2
and 2:1 resonances are $\simeq 10 r_{\rm H}$ and $\simeq 15 r_{\rm H}$, respectively.
This is comparable to the typical 
separation after the oligarchic growth phase: planets are formed
{\it in situ} and then migrate slightly and are captured in the resonances.

As argued above, the formation of resonances between two planets can be discussed in 
terms of two cases of the orbital separation at the time migration begins; namely, 
the planets are well separated from each other $(\Delta > 10 r_{\rm H})$,
or they have relatively close
orbits $(\Delta \simeq 10 r_{\rm H})$. 
The orbital separation of the planets at the onset of migration is characterized by
their migration and growth timescales. 
The growth timescale of a planet $t_{\rm g}$ is determined by the accretion rate of
surrounding planetesimals in the classical planet growth model, which is
similar to or longer than the type I migration timescale at the time of migration
\citep[e.g.,][]{kokubo_ida98,ogihara_ida09}. In this case, the inner planet starts
to migrate before the outer planet does, resulting in an expansion of the orbital separation
$(\Delta > 10 r_{\rm H})$. On the other hand, as planets grow, the surrounding
planetesimals stirred by the planets are fragmented by mutual collisions. The resultant
fragments effectively accrete onto planets \citep{kobayashi10, kobayashi11}; 
$t_{\rm g}$ depends on the initial planetesimal mass
and radial gas density profile, and $t_{\rm g}$ might be shorter than $t_a$ 
in some cases \citep{kobayashi10,kobayashi12} when migration begins. 
If $t_{\rm g} \la t_a$, the two planets can start their migration almost simultaneously,
in which case the orbital separation $(\Delta \simeq 10 r_{\rm H})$ is maintained during migration.
Detailed calculations that include accretion, fragmentation, and migration are needed
to clarify this behavior.

\subsubsection{Systems in Closely Spaced Resonances: Kepler-11 and Kepler-60}
Below we discuss some individual systems. First, in this subsection the systems in 
closely spaced resonances (e.g., 4:3 and 5:4) are considered. As shown in 
Figure~\ref{fig:dis_m1_ta}, the differences in $t_{a,{\rm crit}}$ between adjacent 
pairs are small; for instance, the $t_{a,{\rm crit}}$ difference between the 4:3 and 5:4
resonances is only a factor of two. This means that the migration speeds for
systems in closer resonances can be constrained well.

The Kepler-11 system, in which six planets have been confirmed, has a possible
pair of planets in a 5:4 mean motion resonance (planets b and c).
The critical migration timescales for the 4:3 and 5:4 resonances are $3 \times 10^3 T_{\rm K}$
and $1.5 \times 10^3 T_{\rm K}$, respectively. Thus, if these planets are formed well 
separated from each other $(\Delta > 10 r_{\rm H})$ and then undergo convergent migration, the migration
timescale would be a few thousand times $T_{\rm K}$. This migration speed
is slightly higher than that of type I migration $(C_{\rm I} f_{\rm g} \ga 1)$. 
Because the other planets in the Kepler-11 system are not considered in resonances, it may not be
natural to suppose that the planets undergo significant migration. 
Therefore, it seems likely that they are formed {\it in situ} and exhibit
slight inward migration, which leads to capture into the 5:4 resonance.
In addition, the required high type I migration speed is inconsistent with the 
typical migration speed which is predicted by the population synthesis model
\citep{ida_lin08}, which would also support the \textit{in situ} formation.
The orbital separation is slightly smaller than the typical orbital separation after oligarchic growth
at 1~AU $(\simeq 10 r_{\rm H})$; however, it has been shown that planets can be formed
with smaller orbital separations near the central star 
$(\simeq 7 r_{\rm H})$ \citep{ogihara_ida09}.
It is also possible that the planets formed {\it in situ} and planet b exhibited outward migration
due to the tidal torque from the central star, resulting in capture into the resonance.
Note that the difficulty of {\it in situ} accretion was pointed out in an investigation of the accretion and evolution of
a hydrogen--rich atmosphere for Kepler-11 \citep{ikoma_hori12}.
Another possibility is that the planets are formed in distant orbits and migrate inward
as a resonant convoy, in which the planets can pass through 2:1 resonances
as they are pushed inward by the outer bodies in the resonances \citep{ogihara_ida09}.
However, in such a case, it is likely that other planets are also in resonances
in the final state.

The Kepler-60 system consists of three planets: the inner pair (b and c)
has 5:4 commensurability, and the outer pair (c and d) seems to be in a 4:3 resonance.
If we assume that these planets formed well separated and then migrated inward,
the differential migration timescale would be approximately $5 \times 10^3 - 1\times10^4 T_{\rm K}$.
Using Equation~(\ref{eq:typeI}), we find that the efficiency of type I migration multiplied
by the scaling factor for the gas surface density is $C_{\rm I}f_{\rm g} \simeq 1$.
This means that if the gas surface density is similar to that of the minimum--mass solar
nebula, the migration speed would be that predicted by the linear theory of type I
migration \citep{tanaka_etal02}.
We also do not exclude the possibility of an {\it in situ} formation model for this system.

\subsubsection{Systems with Small Separations: HD~200964, HR~8799, HD~45364, and Gliese~876}
\label{sec:small_separation}
Next, we discuss resonant systems with small orbital separations $(\Delta
\lesssim 2 \sqrt{3} r_{\rm H})$, which can be Hill unstable. 
In the HD~200964 system, two Jovian--mass planets lie in a 4:3 mean
motion resonance (e.g., \citealt{johnson_etal11}; \citealt{wittenmyer_etal12}), and their 
orbital separation is $\simeq 1.5 r_{\rm H}$. 
If we assume that these planets are stable despite their small separation,
the differential migration timescale for passing through the 3:2 resonance
would be extremely short $(\simeq 60 T_{\rm K})$. This migration speed is almost impossible 
to achieve because a large amount of angular momentum should be delivered 
to the disk. Thus, it is not easy to provide a formation model that produces the orbital
properties of the HD~200964 system.
In fact, \citet{rein_etal12} have also claimed that no formation
scenarios are successful in reproducing 4:3 resonant planets similar to those in
the HD~200964 system.
Because the pair of planets around HD~200964 have almost crossing
orbits, they might be in temporary resonance following orbital instability.

The HR~8799 system has inner and outer debris disks \citep{reidemeister09} and four planets that were discovered 
by direct imaging \citep{marois_etal08,marois_etal10}. The masses have not been well 
constrained; the planets have masses between a few and 13 Jupiter masses (e.g., \citealt{marley_etal12}).
Because the orbital separations of the second innermost pair (planets d and c) and 
the outermost pair (planets c and b) are small
$(\simeq 3 r_{\rm H})$, it is plausible that the two pairs of planets are in
2:1 mean motion resonances, which can stabilize the system over the estimated age
of the star \citep{reidemeister09,fabrycky_murray-clay10}.
In this case, our model gives a lower limit on the migration timescale of $40 T_{\rm K}$.
This value is quite small, so it may not provide a useful constraint. 
In addition to this lower limit, the time of passage of the Hill unstable region should be shorter
than the orbital unstable time for a system with small orbital separation 
$(\Delta \lesssim 2\sqrt{3}~r_{\rm H})$.
According to \citet{fabrycky_murray-clay10}, the crossing time, which is the timescale of the
initiation of orbital instability, is $\simeq 3 \times 10^5~{\rm yr} = 3 \times 10^3 T_{\rm K}$; 
therefore, the actual migration timescale would be between 
$40~T_{\rm K}$ and $3 \times 10^3 T_{\rm K}$.
The type II migration expected for such massive planets cannot bring
about the short migration timescale. 
\citet{reidemeister09} pointed out that if the masses of the planets are small,
the system can be Hill stable.

Gliese~876, an M--dwarf star with a mass of $0.33 M_\odot$, harbors four planets: 
planets c and b have comparable masses, and planets d and e are relatively
small, with masses of $\sim 10 M_\oplus$. Planets c--b and b--e are in or close to
the 2:1 resonance.  The orbital separations for the inner pair (c--b) and the outer pair (b--e)
are $3.2 r_{\rm H}$ and $3.5 r_{\rm H}$, respectively. 
These planets are believed to be in the 2:1 resonance to stabilize their orbits
at such a small orbital separation (e.g., \citealt{marcy_etal01}).
The lower limits to the differential migration timescales are $40 T_{\rm K}$ for the
inner pair and $400 T_{\rm K}$ for the outer pair, which is consistent
with the migration timescale obtained from
a hydrodynamical simulation of this system by \citet{kley_etal04}.

Finally in this subsection, we add some comments on planets with small separations
that are still Hill stable.
The HD~45364 system consists of two planets with masses comparable to those
of Saturn and Jupiter \citep{correia_etal09}, which seem to be in a 3:2 
mean motion resonance with a separation of $3.9 r_{\rm H}$. This system 
should have undergone rapid migration to pass through the 2:1 resonance. 
The critical migration timescale for
capture into the 2:1 resonance is given by $t_{a,{\rm crit}} \simeq 10^3 T_{\rm K}$;
thus, the differential migration timescale should be at least shorter than $10^3 T_{\rm K}$.
\citet{rein_etal10} also examined this system both numerically and analytically,
and proposed that a relative migration timescale shorter
than $800 T_{\rm K}$ is needed to pass through the 2:1 resonance, which is 
consistent with our results.
Our model also provides a lower limit of $200 T_{\rm K}$ on the migration timescale.
The planets in this system would have undergone type III migration.

\subsection{Other Systems}
Here we apply our results to resonant systems in the solar system. 
One good example is the Galilean satellites around Jupiter, where 
Io--Europa and Europa--Ganymede are in the 2:1 mean motion resonance.
Replacing the stellar mass $(M_*)$ with the Jupiter mass, our model gives a 
lower limit of $1 \times 10^4 T_{\rm K}$ on the differential migration speed.
This constraint is consistent with the result of \textit{N}--body work on the
formation of the Galilean satellites (Equation~[32] in \citealt{ogihara_ida12}).

Some trans--Neptunian objects are in mean motion resonance with
Neptune. The population of objects in the 3:2 resonance is much higher than
that in the 2:1 resonance. 
The objects were captured into the mean motion resonances during outward migration of
Neptune. Simulations by \citet{ida_etal00} showed that the
tendency could be explained by the migration speed of Neptune. 
We apply our formula for the inward migration of the outer objects 
to resonant capture by an outwardly migrating Neptune.  
The obtained migration timescale of $3$--$10$\,Myr is plausible for 
the high population of the 3:2 resonance, which is roughly consistent with \citet{ida_etal00}.
Furthermore, this estimated migration timescale is also consistent with that obtained 
by \citet{murray-clay_chiang05} of $1$--$10$\,Myr, which is derived from studying the 
proportion of TNOs in the leading and trailing islands of the 2:1 resonance.
Neptune's outward migration, which is caused by interaction with the surrounding
planetesimals, and other planets may not be as smooth as
we assume \citep{levison_etal08}. The application of our model to these
objects should be done carefully.

We can further discuss other planet formation models, which assume 
the establishment of mean motion resonances.
In the Grand Tack model \citep{walsh_etal11}, Saturn migrates
faster than Jupiter, which results in capture into a 3:2 mean motion resonance
with Jupiter. For capture into the 3:2 resonance, the differential
migration timescale between the two planets should be $100$--$500 T_{\rm K}$,
which is smaller than the typical type I or II migration timescales.
\citet{walsh_etal11} considered high--speed type III
migration, which is necessary to realize the Grand Tack scenario.

\section{CONCLUSION}
\label{sec:conclusions}
We investigated capture into first--order mean motion resonances in a system
of two bodies undergoing damping of the eccentricity and semimajor axis using
\textit{N}--body integrations. 
In some of our calculations, we considered the case in which the mass of one body is 
negligible. In addition, we also studied systems with equal masses.
In fact, orbital calculations were performed with a wide range of parameters; we found 
that the critical migration timescale can be described using the mass ratio between
the larger body and the central object, and depends weakly on the $e$--damping
timescale and initial eccentricity. The empirical formula is given by 
Equation~(\ref{eq:ta_crit}), where the critical migration timescale for equal--mass 
bodies is about an order of magnitude shorter than that for systems with a massless
particle. We also confirmed the power--law dependence of the mass with index $-4/3$.
This dependence is also supported by analytical arguments that compare
the resonant libration timescale and the migration timescale. 
Additional simulations of closely spaced resonances were run, and empirical
fits to the results were derived. All the fitting formulae from our calculations 
are shown in Figure~\ref{fig:dis2_m1_ta} and Table~\ref{tbl:fitting}. 

The empirical formula we derived can constrain the relative migration speed 
in systems of two bodies undergoing convergent migration toward capture into 
mean motion resonances. This means that our model can be useful for understanding
the origins of exoplanet systems in resonances. 
For systems in closely spaced mean motion resonances (e.g., Kepler-11, Kepler-60), the
migration timescale can be well constrained. It is also possible that the planets formed {\it in situ}.
The systems in which the orbital separation is smaller than $2 \sqrt{3} r_{\rm H}$
(e.g., HR~8799, Gliese~876) are believed to become stable owing to resonant
effects. Lower limits to the relative migration timescale were placed on several systems
in 2:1 resonances. The origin of the HD~200964 system, which is in 4:3 resonance, 
remains unclear. Our model also provides constraints on the migration timescale
of systems other than exoplanets (e.g., the Grand Tack model of the solar system).
Furthermore, when the number of discovered exoplanets in resonances increases 
sufficiently in the future, the typical type I and II migration timescales can be obtained using
our results; future observations will allow us to tackle this issue.
Note that if the eccentricity at a resonant encounter is large, the
capture into resonance becomes probabilistic. In this case, our model can 
provide only a necessary condition.

In this work, we considered first--order mean motion resonances, which are 
certainly important for planets both inside and outside the solar system. 
In addition, higher--order mean motion resonances (e.g., 3:1 and 5:2) can also be important for
specific systems (e.g., \citealt{michtchenko_ferraz-mello01}; \citealt{steffen13}); therefore, it would
be worth examining the conditions for capture into these resonances in future work.

\subsection*{ACKNOWLEDGMENT}
We thank to an anonymous referee for detailed helpful comments.
We also thank S.~Inutsuka for helpful discussions. 
The numerical computations were conducted in part on 
the general--purpose PC farm at the Center for Computational Astrophysics, 
CfCA, of National Astronomical Observatory of Japan.
This work was supported by a Grant--in--Aid for JSPS Fellows (23004841).
HK gratefully acknowledges the support of a Grant--in--Aid from MEXT
(23103005).

{}

\begin{deluxetable}{lllllll}
\tablecolumns{8}
\tablewidth{0pc}
\tablecaption{Systems in or near first-order mean motion resonances}
\startdata
\hline \hline
Star		& $M_* (M_\odot)$	& Planet	& $M (M_\oplus)$	& $a$ (AU)& MMR [pair]	& Period ratio\\
\hline
Gliese 876	& 0.334	& d	& 6.68			& 0.0208	&		& 		\\
			& 	& c	& 227			& 0.130	& 2:1 [b]	& 2.03	\\
			& 	& b	& 723			& 0.208	& 2:1 [e]	& 2.03	\\
			& 	& e	& 14.6			& 0.334	&		& 		\\
HD 37124		& 0.83	& b	& 215			& 0.534	& 		& 		\\
			& 	& c	& 207			& 1.71	& 2:1 [d]	& 2.10	\\
			& 	& d	& 221			& 2.81	&		& 		\\
HD 73526		& 1.08	& b	& 922			& 0.66	& 2:1 [c]	& 2.00	\\
			& 	& c	& 795			& 1.05	&		& 		\\
HD 82943		& 1.18	& c	& 639			& 0.746	& 2:1 [b]	& 2.01 	\\
			& 	& b	& 556			& 1.19	& 		& 		\\
HD 128311	& 0.84	& b	& 693			& 1.10	& 2:1 [c]	& 2.05	\\
			& 	& c	& 1020			& 1.76	&		& 		\\
HR 8799		& 1.56	& e	& 2861			& 15		& 		& 		\\
			& 	& d	& 3179			& 27 	& 2:1 [c]	& 2.00	\\
			& 	& c	& 3179			& 43		& 2:1 [b]	& 1.99	\\			
			& 	& b	& 2225			& 68		&  		& 		\\
Kepler-9		& 1	& d	& 6.99			& 0.0273	& 		& 		\\
			& 	& b	& 80.1			& 0.14	& 2:1 [c]	& 2.02	\\
			& 	& c	& 54.4			& 0.225	& 		& 		\\
Kepler-18		& 0.972	& b	& 6.90			& 0.0447	& 		& 		\\
			& 	& c	& 17.2			& 0.0752	& 2:1 [d]	& 1.94	\\
			& 	& d	& 16.5			& 0.117	& 		& 		\\
Kepler-48		& 0.89	& b	& 4.79			& 		& 2:1 [c]	& 2.02	\\
			& 	& c	& 10.6			& 		& 		& 		\\
Kepler-51		& 1.00	& b	& 55.9			& 		& 2:1 [c]	& 1.89	\\
			& 	& c	& 36.2			& 		& 		& 		\\
Kepler-52		& 0.54	& b	& 4.61			& 		& 2:1 [c]	& 2.08	\\
			& 	& c	& 3.51			& 		& 		& 		\\
Kepler-53		& 0.98	& b	& 8.90			& 		& 2:1 [c]	& 2.07	\\
			& 	& c	& 10.8			& 		& 		& 		\\
Kepler-56		& 1.37	& b	& 16.0			& 		& 2:1 [c]	& 2.04	\\
			& 	& c	& 69.7			& 		& 		& 		\\
Kepler-57		& 0.83	& b	& 5.03			& 		& 2:1 [c]	& 2.03	\\
			& 	& c	& 2.47			& 		& 		& 		\\					
mu Ara 		& 1.08	& c	& 10.6			& 0.0909	& 		& 		\\
			& 	& d	& 166			& 0.921	& 2:1 [b]	& 2.07	\\
			& 	& b	& 533			& 1.5	& 		& 		\\
			& 	& e	& 577			& 5.235	&		& 		\\
24 Sex		& 1.54	& b	& 633			& 1.33	& 2:1 [c] 	& 1.95	\\
			& 	& c	& 273			& 2.08	& 		&		\\
HD 45364		& 0.82	& b	& 59.5			& 0.681	& 3:2 [c]	& 1.51	\\
			& 	& c	& 209			& 0.897	&		& 		\\
Kepler-49		& 0.55	& b	& 7.86			& 		& 3:2 [c]	& 1.51	\\
			& 	& c	& 6.88			& 		& 		& 		\\
Kepler-54		& 0.51	& b	& 4.61			& 		& 3:2 [c]	& 1.51	\\
			& 	& c	& 1.53			& 		& 		& 		\\
Kepler-55		& 0.62	& b	& 6.23			& 		& 3:2 [c]	& 1.51	\\
			& 	& c	& 5.12			& 		& 		& 		\\
Kepler-58		& 0.95	& b	& 8.22			& 		& 3:2 [c]	& 1.52	\\
			& 	& c	& 8.71			& 		& 		& 		\\
Kepler-59		& 1.04	& b	& 1.22			& 		& 3:2 [c]	& 1.51	\\
			& 	& c	& 4.08			& 		& 		& 		\\					
HD 200964	& 1.44	& b	& 588			& 1.60	& 4:3 [c]	& 1.34	\\
			& 	& c	& 286			& 1.95	&		& 		\\
Kepler-60		& 1.11	& b	& 5.46			& 		& 5:4 [c]	& 1.25	\\
			& 	& c	& 6.44			& 		& 4:3 [d]	& 1.33	\\
			& 	& d	& 6.88			& 		& 		& 		\\
Kepler-11		& 0.961	& b	& 4.30			& 0.091	& 5:4 [c]	& 1.26	\\
			& 	& c	& 13.5			& 0.106	& 		& 		\\
			& 	& d	& 6.10			& 0.159	& 		& 		\\			
			& 	& e	& 8.40			& 0.194	& 		& 		\\
			& 	& f	& 2.30			& 0.25	& 		& 		\\
			& 	& g	& 302			& 0.462	& 		& 		\\
Kepler-50		& 1.24	& b	& 5.07			& 0.077	& 6:5 [c]	& 1.20	\\
			& 	& c	& 8.28			& 0.087	& 		& 		
\enddata
\tablecomments{Values are taken from the Open Exoplanet Catalogue. Shown 
(left to right) are the Star, the stellar mass, the identifier of the planet, the planet mass,
the semimajor axis, the commensurability of the resonant pair, and the period ratio of
the resonant pair.
}
\label{tbl:exoplanets}
\end{deluxetable}
\end{document}